# URBAN HOUSEHOLD BEHAVIOR IN INDONESIA: DRIVERS OF ZERO WASTE PARTICIPATION

*Faizal Amir\*, Alimuddin S.Miru, Edy Sabara*

*Department of Population and Environmental Education, Universitas Negeri Makassar, Indonesia, 90222*

\*Corresponding author: faizalamir64@unm.ac.id

**Abstarct**

The 3R-based Zero Waste approach minimizes household solid waste through the Reduce, Reuse, and Recycle principles. This study explores the relationship between household solid waste environmental knowledge, personal attitude, subjective norms, and perceived behavioral control as critical behavioral predictors. The researchers used a structured survey of 1,200 urban households across 12 Indonesian cities. The data collected were analyzed using Pearson correlation and multiple regression analysis. The findings showed that perceived behavioral control remains the strongest predictor of household waste management behavior ($\beta = 0.367$, $p \leq 0.001$), followed by subjective norms ($\beta = 0.358$, $p \leq 0.001$) and environmental knowledge ($\beta = 0.126$, $p \leq 0.001$). It indicates that individuals have confidence in handling waste in their practical behavior. Overall, perceived behavioral control, subjective norms, and environmental knowledge influence a household's behavior toward Zero Waste. Because households generate and dispose of waste regularly, they form the backbone of overall municipal waste management programs in any country. These findings provide valuable insights for enhancing behavioral interventions and informing policy design based on the Theory of Planned Behavior (TPB).

## 1. Introduction

Disposing of household waste is one of the most critical environmental issues directly related to urban sustainability and climate change mitigation. The 3Rs-based Zero Waste approach has evolved into a holistic approach for reducing waste generation and enhancing resource utilization. Proper Zero Waste projects can significantly reduce reliance on landfills, decrease greenhouse gas (GHG) emissions, and encourage responsible use of natural resources [1,2]. Nevertheless, while public awareness has been growing and regulations have been advancing, the translation of policy intentions into household-level commitment to sustainable waste practices lags [3].

Nowhere is this gap starker than in Indonesia, where a combination of high levels of urbanization and consumerism has led to an unprecedented rise in municipal solid waste. According to national estimates, waste generation has increased from 29.3 million tons in 2019 to around 34.9 million tons in 2022 [4,5]. Although policy efforts aim for sustainable waste management, household participation in waste sorting and recycling remains low [6,7]. The lack of infrastructure and underdeveloped urban systems further exacerbate the issue of developing pro-environmental behaviors at the household level [8,9]. Based on the Theory of Planned Behavior (TPB), individuals' household waste management behavior is determined by their attitudes, subjective norms, and perceived behavior control [10]. Although the level of environmental consciousness at the household is high, it is characterized by relatively low perceived efficacy and enabling conditions, which results in low conversion to environmentally conscious behavior [11,12,13,14]. Social mobilization and community-based programs have also been successful in some settings; however, structural obstacles and lack of information remain formidable challenges [15,16,17].

Recent literature [18,19,20,21] identifies several gaps that have frustrated the efforts of existing interventions. Most research employs cross-sectional designs, which do not provide insight into longitudinal behavioral characteristics [22,23]. Additionally, only a few studies directly address regional and socio-economic diversity, essential to understanding the urban waste behavior of a heterogeneous country like Indonesia [24, 25, 26]. Crucially, the common perception that a linear correlation exists between people's environmental concerns and their environmentally responsible actions is overly simplistic. Prior literature has undervalued the importance of perceived behavioral control, specifically the disbelief in one's capacity to escape the restraint of infrastructure and institutional constraints to perform sustainable behaviors [27,28,29].

Subjective norms also warrant further investigation. Even in environmentally literate audiences, behavior may be challenging to motivate when apparent peer acceptance is not evident. Cultural factors and collective behaviors significantly influence results, especially in communalistic societies where adherence to social norms is a primary motivator [25, 26]. Additionally, differences in the availability of city support services, technology infrastructure, and outreach programs contribute to spatial differences in waste behavior responses [27].

Demographic elements, such as gender, age, and education, influence how households receive, consume, and translate environmental information into action. For instance, women typically organize domestic practices, making them central actors in sanitary consumer familiarity decisions. Similarly, digital campaigns may be more successful among younger adults [28]. A contextually sensitive understanding of these demographic and cultural dimensions is a critical prerequisite for developing contextually specific interventions. This study employs an integrated analytical model incorporating psychological, social, and contextual factors driving households' solid waste management to address these gaps. Drawing on a multi-city dataset of 1,200 households in twelve Indonesian cities, the research examines the influence of environmental knowledge, attitudes, subjective norms, and perceived behavioral control on household behavior in Zero Waste.

This research assumes that environmental knowledge, subjective norms, and perceived behavioral control are significant influencing factors for household solid waste management practices in urban areas of Indonesia, with perceived behavioral control being the most dominant predictor.

Through these theoretical, empirical, and practical considerations, this study offers new insights into the broader debate surrounding sustainable urbanism. It aims to inform decision-makers and urban planners by identifying key behavioral levers for effective community-based waste management interventions.

## 2. Materials and Methods
## 2.1 Materials
### 2.1.1 Zero Waste (ZW) and domestic waste management.
The ZW movement has evolved as an international approach to minimize waste and enable a closed-loop in resource use by adhering to the 3R (reuse, recycle, reduce) principles. The ultimate goal is to work toward zero waste creation and restrict the remaining so that it can be reused or recycled with little environmental impact [29]. Households, which are overwhelmingly located in urban areas of developing countries, play a significant role in generating municipal solid waste (MSW) [30].
Behavioral change has become central to the success of ZW programs at both individual and societal levels. Public engagement and behavior change are necessary to complement successful policies and infrastructure, effectively translating policy into action [31].

## 2. 1. 2. Theory of Planned Behavior (TPB)

The theoretical framework used for this study was the Theory of Planned Behavior (TPB), developed by Ajzen. This framework posits that behavioral intention is determined by attitude toward the behavior, subjective norms, and perceived behavioral control [32]. These factors drive a person's propensity to engage in particular behaviors.

There is empirical evidence to support the application of the TPB to waste-related research. For instance, [33] demonstrated that perceived behavioral control was a significant predictor of household waste recycling in China and significantly predicted waste recycling in Chinese households. Furthermore, community norms and peer pressure substantially influenced pro-environmental behavior in South Korea [34].

## 2. 1. 3. Environmental Knowledge and Attitudes

The knowledge base is a key element in realizing responsible waste behavior. Individuals with a higher level of understanding about the impacts of improper waste management are also more likely to participate in recycling programs and waste reduction activities [35]. However, knowledge alone does not necessarily lead to behavior change. Environmental consciousness can be heightened through environmental education, while further motivational incentives are necessary to foster action [36].

Attitude factors, individual appraisals of a behavior's positive or negative aspects, are also involved. Favorable attitudes toward recycling correlate with increased recycling behavior [37]. However, attitudinal factors alone are insufficient without facilitating beliefs or societal support [38].

## 2. 1. 4. Subjective Norm and Social Influence

Subjective norm is defined as the perception of social expectation about behavior. According to [39], households are surrounded by societal norms that support recycling and are more likely to engage in this behavior themselves [38, 40]. Social learning and community-based interventions are effective in reinforcing pro-environmental norms.

[41] highlights successful community-driven efforts in South Korea, where local leaders played a crucial role in promoting the cooperative management of solid waste. These efforts underscore the opportunity to mobilize peer influence and community-based networks.

## 2. 1. 5. Perception of Behavior Control and Constraints.

Perception of Behavior Control (PBC) reflects an individual's judgment about how easy or difficult it is to perform a particular behavior. This includes internal factors like confidence and external resources like infrastructure [42]. Research has validated PBC for its predictive value in waste behaviour. [43] reported that a recycling facility encourages waste segregation within homes, whereas the lack of one discourages recycling.

Demographic variables can refine perceived control. [44] found that younger and more educated respondents expressed greater confidence in handling waste, suggesting a need for targeted interventions based on demographics

Although widely used, the TPB method has limitations. Cross-sectional studies predominate, which restricts the ability to monitor changes in behavior over time [45]. Furthermore, regional and socioeconomic disparities have been underinvestigated, particularly in developing country contexts, such as Indonesia [46, 47].

In addition, environmental education is not a universal cure-all as often touted; the WHAT and HOW of environmental education matter. [48] Do draw attention to the need to advocate participatory and localised educational modalities in which households are directly involved.

Indonesia's context is characterized by its diversity in geography, rapid urbanization, and economic development disparities. National policies like the ZW campaign have established preliminary frameworks; however, community engagement remains inconsistent [49].

[7] highlighted the role of social norm expectations and local leadership in influencing household waste behavior, and [24] found significant regional differences in recycling behavior as a function of access to infrastructure. Most prior work conducted in Indonesia has framed its research regarding attitudes and knowledge, without a corresponding consideration of perceived behavioral control, which this study aims to address across multiple cities.

This review provides evidence to support the assertion that the TPB is an appropriate model for gaining insight into domestic waste behavior and for acknowledging the value of incorporating attitudes, norms, perceived control, and enabling environments in developing sustainable household waste behavior.

## 2. 2. Method

### 2. 2. 1. Research Design

The current study is a cross-sectional survey research that utilizes routinely collected data to analyze the psychological and socio-contextual determinants of household solid-waste management (HSWM) behavior under the ZW concept. Specifically, the study employed the TPB model to investigate how environmental knowledge, attitude, subjective norms, and perceived behavioral control are associated with behavioral intentions and behaviors among urban households.

### 2. 2. 2. Study Area and Sampling

This study's cities were purposively selected to represent various geographical features, population densities, and waste management infrastructure. The chosen cities were Padang, DKI Jakarta, Bandung, Yogyakarta, Surabaya, Makassar, Manado, Banjarmasin, Denpasar, Ternate, Bima, and Sorong. Stratified random sampling was employed, with 100 heads of households included in each city, resulting in 1,200 valid samples. Sampling was stratified according to demographic (i.e., age, education, income) and municipal (i.e., city size, local policies) variables. Housewives were targeted because they are often the primary decision-makers in household matters, including waste management.

### 2. 2. 3. Measurements and Data collected

The instrument used to get the data is a questionnaire. The questionnaire measured four key psychological constructs derived from the TPB. Environmental knowledge was assessed to ascertain the respondents' awareness of waste issues and sustainability actions. Attitudes were the individual appraisal of waste sorting and recycling behavior. The perceived social pressure to perform or not to perform sustainable waste-related actions was included as the subjective norms, and belief in capacity was included in the perceived behavioral control, which assessed the respondents' confidence regarding effectively managing waste.

The questionnaire consisted of five sections, and data were gathered with a self-administered structured questionnaire:

Demographics – Age, sex, education, income, size of household.

Knowledge of the Environment – 6 items (e.g., "I know the environment words to stop at the waste, inappropriate waste, where is it harmful"), α = 0.81.
Waste Behavior Attitude – 5 items (e.g., "Recycling is good for my neighborhood"), α = 0.83.
Subjective Norms – 5 items (e.g., "My family wants me to separate waste at home"), α = 0.79.
Perceived Behavioral Control (PBC) – 6 items (e.g., "I have the resources needed to handle my household waste properly").84.
All measures used a 5-point Likert scale anchored at 1 (Strongly Disagree) and 5 (Strongly Agree). The items were taken from proven Theory of Planned Behavior (TPB) instruments and contextualized in the urban Indonesian, following the procedure outlined by [50].

### 2. 2. 4. Theoretical Framework

According to the TPB, behavior intention is determined by the attitude, subjective norm, and perceived behavior control. The model was operationalized as depicted in Figures 1 and 2, where household behavior was predicted as a function of the psychological variables analyzed.

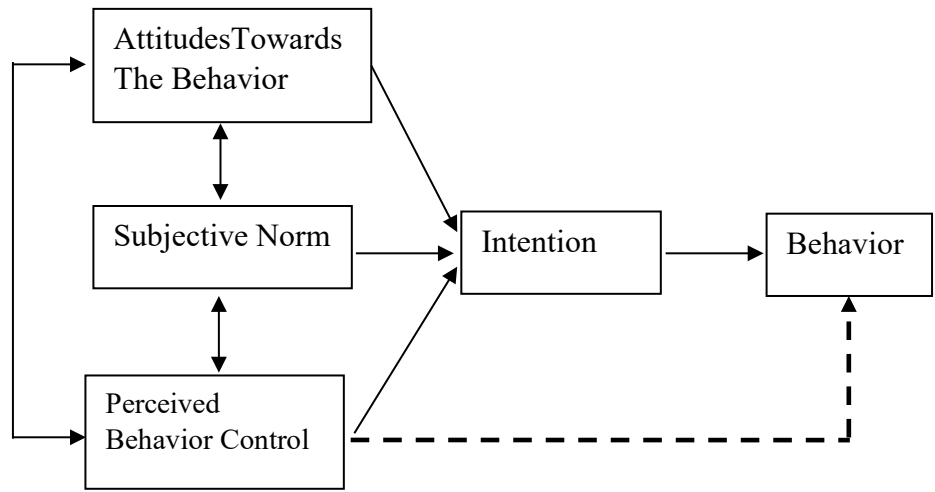

**Figure 1**: Planned Behavior Theory (Source: Ajzen, 1991

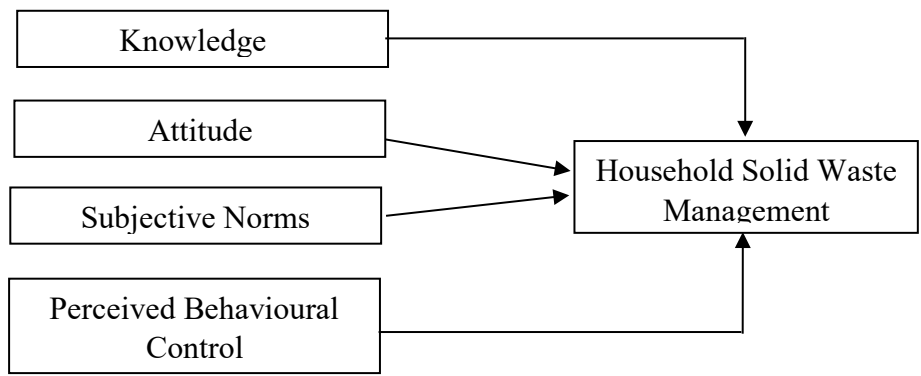

**Figure 2**: Research framework based on Ajzen's TPB (1991)

Data were processed through SPSS. Demographic data were summarised using descriptive statistics. A Pearson correlation analysis examined the relationship between TPB variables and waste behavior. Multiple linear regression was performed to investigate the predictive environmental knowledge, attitude, subjective norms, and PBC toward household waste behavior, at p < 0.05. All effect sizes are presented as standardized beta coefficients (β).

## 4. Result
### 4.1. Environmental Knowledge

Knowledge of the environment was high based on data obtained from 1,200 urban households in 12 cities in Indonesia. Specifically, 97.6% of respondents agreed that buying in bulk helps reduce waste, and 98.5% agreed that having a list when shopping minimizes packaging. In addition, 93.8% knew the 3R principle (Reduce, Reuse, Recycle). The average score in the environmental knowledge test was 17.28, indicating a high exposure to ecological sustainability ideas.

However, 35.4% of those surveyed knew of the future 2024 waste sorting law. This indicates a lack of understanding of government policy, despite a strong fundamental knowledge of waste reduction practices (Figure 3).

Figure 1 illustrates that the heads of households were highly aware of the benefits of the 3R approach.

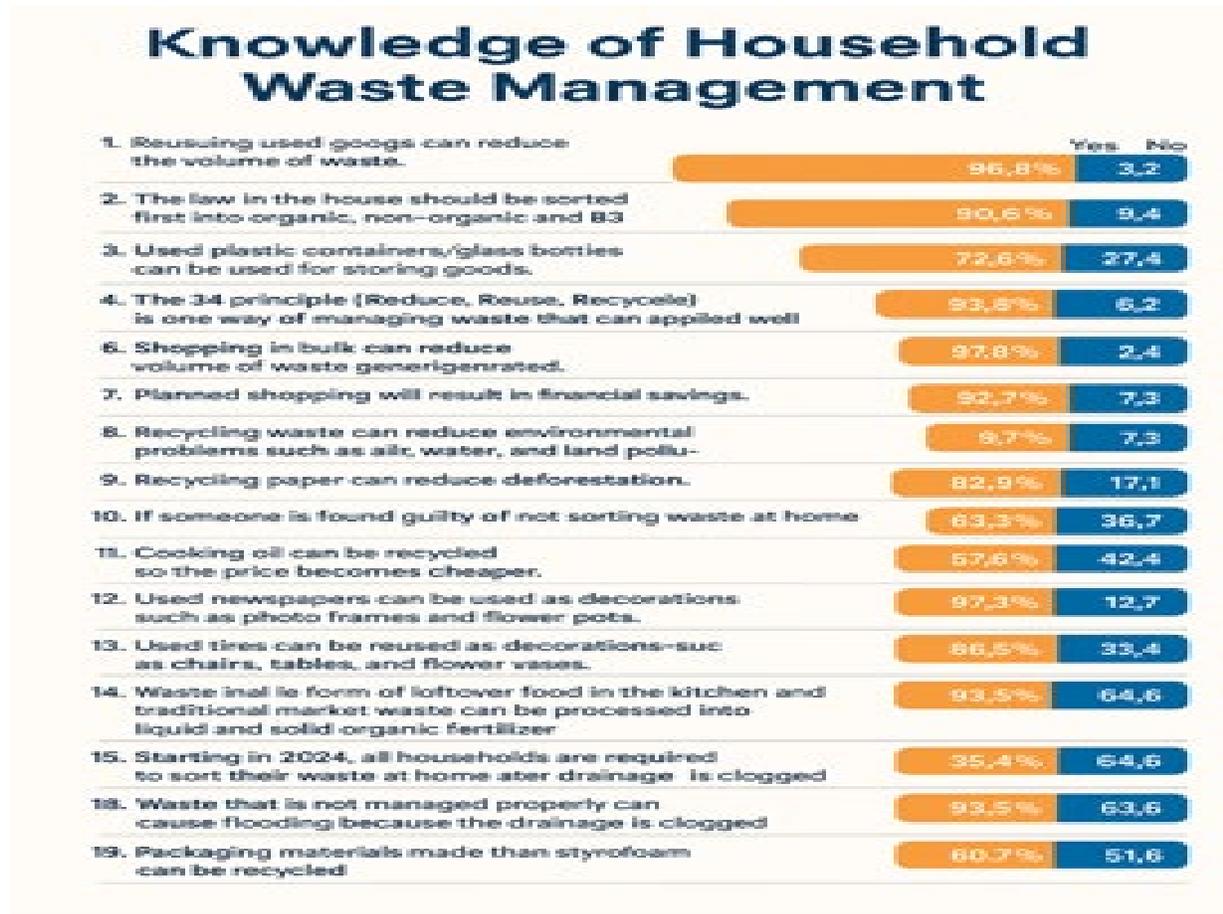

**Figure 3.** Individual's Environmental Knowledge

## 4.2. Attitude toward Household Waste Management

The survey findings revealed that 73% of people believe good waste management leads to a healthier environment. This indicates a general willingness to support climate action among urban households. Furthermore, more than 70% agreed or strongly agreed with various statements for waste minimization, reuse, and recycling.

Yet, about half of the respondents identified waste management as a hassle, due to reasons such as infrastructure problems or inconvenience. A minority of the respondents, 34.5% (agree, Figure 4), believe in the Importance of Waste management and collective effort. There is a mixed perception of social and personal responsibility, with only a third of the respondents strongly agreeing that waste disposal is a collective duty.

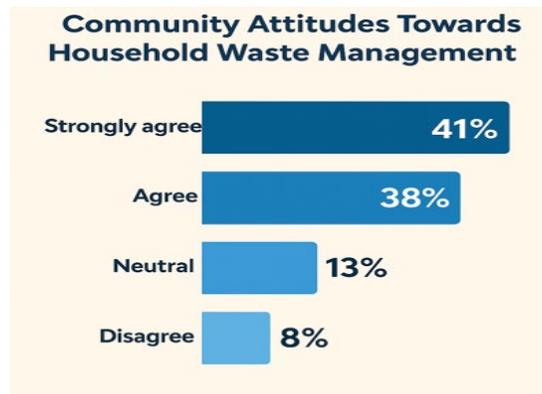

**Figure 4.** Community Attitudes

## 4.3. Perceived Behavioral Control

The diagram in Figure 5 presents respondents' perceived behavior control (PBC) factors regarding household waste management behaviors. Indeed, the fact that 77% of respondents agreed that "everyone is responsible" for managing household waste would indicate a high level of social responsibility being perceived. At the same time, however, about 70% also agreed with the statement that waste separation is "tiring" or "difficult," indicating that perceived exertion continues to pose a formidable obstacle.

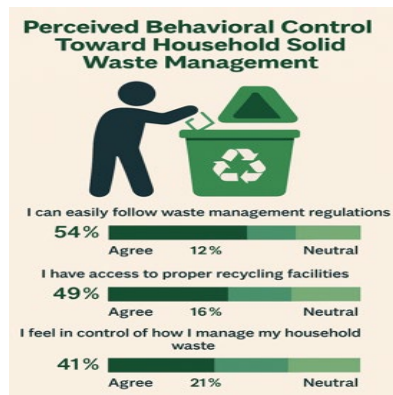

Figure 5. Perceive Behavior Control

These divergent trends are indicative of a multidimensional psychological-behavioral profile. Many people there express a sense of their responsibility and willingness to participate, but at the same time, they see efficient limitations. The gap between moral responsibility and the possibilities of action for behavior is evident. It has been suggested that this gap echoes a gap in the research agenda that must be more fully addressed if we are to grasp how infrastructural and emotional burdens inform environmental practices.

### 4.4. Subjective Norms

The distribution of perceived subjective norms regarding household waste is shown in Figure 6 for Indonesian urban respondents. It indicates that 70.9% of respondents perceived that family members were the most influential people affecting their waste behavior, followed by neighbors, community leaders, and peers. Furthermore, 72.6% noticed the environmentally friendly behaviour of the people in their immediate area, like waste separation, recycling, etc.

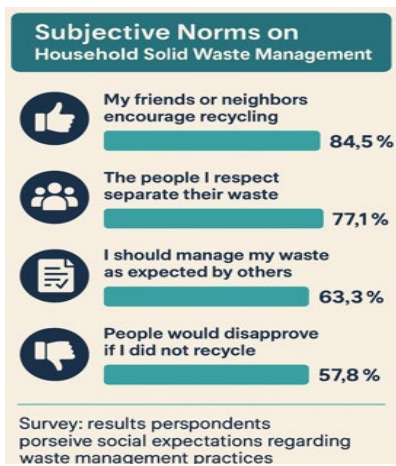

**Figure 6**. Subjective Norm

These results imply that family expectations and household models are most influential in shaping individual household waste behaviors. Nevertheless, the relative lack of community-level actors visible suggests an overlooked source of broader group mobilisation.

### 4.5. Household Waste Management Behaviors

Household waste management practices among 1,200 urban respondents from various regions in Indonesia are presented in Figure 7. The results show that more than 70% of the households commonly practise at least one sustainable waste management method (e.g., reduce, reuse, and recycle). More specifically, 77.9% of the respondents reported that saleable recyclables from households were sold (e.g., plastic bottles, metal, paper), and around 80% reused containers or paper materials within their families.

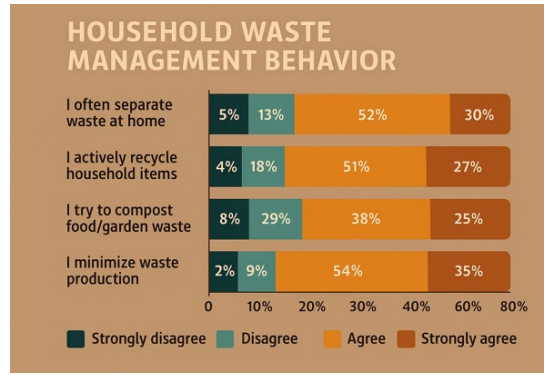

**Figure 7**. Household Waste Management Behavior

However, despite these promising statistics, the figures also reveal that single-use items are still being used, suggesting that the behavioral change is not yet complete. The coexistence of these sustainable behaviors and waste-generating routines indicates a behavioral paradox.

**4.6. Relationship between psychological factors and waste behaviour**

Those who were not familiar with that schedule (60.7%) did not know the day waste would be collected. 63.9% of those who knew the day waste was going to be collected sometimes gave their waste to children who made waste collection, and only 6% did so.

The correlation coefficients of psychological variables with household waste behavior are given in Figure 8. The findings indicate that all constructs of the Theory of Planned Behavior (TPB) are positively related to waste behavior. PBC was identified as the strongest predictor (r = 0.391), followed by environmental knowledge (r = 0.172), subjective norms (r = 0.162), and attitudes (r = 0.143).

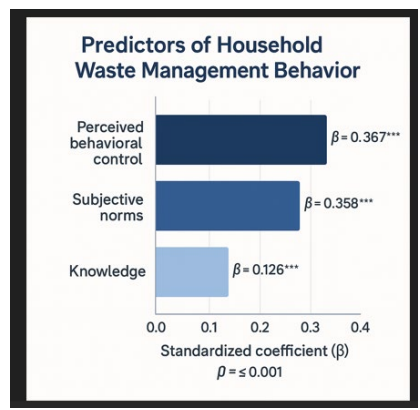

**Figure 8.** Predictors of Household Waste Management Behavior

These statistical relationships suggest that respondents are more likely to perceive themselves as capable of conducting the behavior when they feel they know more about it, are more supported in performing the behavior, and believe waste should be managed socially. Furthermore, this perception is more likely to lead to the adoption of the behavior. Indeed, the strong effect of PBC, in particular, underscores the role of self-belief and free will in environmental action.

### 4.7. Analysis of the Influencing Factors of the Regression Model

Results from multiple linear regression analysis suggest that, among the psychological variables measured, PBC is the most significant predictor of household waste management behavior, with a β of 0.367 and p ≤ 0.001. Next, subjective norms (β = 0.358, p ≤ 0.001) and environmental knowledge (β = 0.126, p ≤ 0.001) follow.

All these coefficients indicate that waste sorting is most likely to occur when the household is confident in having the necessary abilities to execute PBC, feels pressure to act, or feels support (subjective norms), and is aware of the environmental consequences of waste (knowledge). Each of the predictors is statistically significant, thereby confirming their importance in the model of behavior through the TPB.

## 5. Discussion

### 5.1. Knowledge of Household Waste Management

The environmental literacy of respondents, characterized by extremely high exposure to 3R and a habit of minimizing trash behaviors (more than 90%), suggests that urban Indonesian households possess basic knowledge of the environment. Our results extend the evidence in the literature, which indicates that cognition serves as a cognitive antecedent of pro-environmental behavior [51, 52, 53]. Yet, the relatively low levels of awareness of government-initiated behaviors (35.4%, for example, regarding the 2024 waste sorting regulation) indicate a substantive policy communication gap [54,55].

Importantly, this right highlights the precariousness of assuming a linear relationship between information and action. According to the Theory of Planned Behavior (TPB) [10], knowledge is implicitly included as an antecedent to attitudinal and normative beliefs. However, TPB can overlook the consideration of how contextual influences, such as structure, institutions, and emotions, moderate the transformation of knowledge into action. [56,57] observed, "knowledge without a platform and motivation can be inert."

This study provides empirical support for the notion of a "knowledge-action gap," as knowing what to do does not mean it will be done. The policy implication here is important: investing in the provision of information in isolation, such as pamphlets, seminars, or media campaigns, could reach a saturation point without being coupled with improvements in accessibility, trust, or perceived levels of control. [58] reported policy inaction despite a high level of exposure to information.

Furthermore, the belief that more information is always better can also lead to "information fatigue," as citizens become disempowered not by ignorance but by the absence of clear pathways to meaningful action. This problem is widespread, especially in lower-middle-income urban areas, where economic or logistical conditions limit the opportunity to act on knowledge [59, 60].

Here, the key lesson is that knowledge is not an end in itself, but rather a means to an end, only effective when mobilized in systems of empowerment. For example, community-based waste programmes with active role models, real-time feedback (e.g., sorted waste audits), and participative planning can convert passive awareness into collective efficacy [61, 62, 63].

In addition, qualitative aspects of knowledge, such as moral relevance, emotional salience, and local applicability, are often overlooked in structured questionnaires. This curtails the extent to which current assessment forms can capture the full potential of knowledge as a transformative resource. In this respect, further research should also have more ethnographic and/or participatory dimensions that examine how individuals internalize, perceive, and value environmental knowledge in everyday contexts.

Finally, while education is a crucial long-term instrument, it must come in a new guise. Environmental literacy as part of the curriculum should be integrated with local and experiential curricula that connect actions to observable effects. This is particularly important for children, as early experiences are the best predictors of adult environmental commitment [64].

Although environmental knowledge is prevalent in Indonesian urban societies, its behavioral relevance is conditional. Using a critical lens, one might argue that knowledge can be performative rather than transformational, stripped of supportive settings, infrastructure, and effective appealing messaging. There is a need for future interventions to embed understanding as part of a larger ecology of behavior change; a system that is sensitive to real-life limitations, cultural scripts, and a system of power.

## 5.2. Attitude toward Household Waste Management

A large percentage of respondents (73%) agree that sound waste management is good for the environment, reiterating the positive attitude trend in previous urban environmental behavior studies [65]. However, the simultaneous discovery that almost 50% of these respondents consider waste management painful suggests an enduring attitude-behavior discrepancy. This duality expresses a critical disconnect: although knowledge and awareness of the importance of waste management are present, putting them into practice is impeded by any perceived degree of effort, inconvenience, or infrastructure constraint [66].

This discordance is reminiscent of a phenomenon often observed in the behavior change literature, the gap between beliefs and practices, otherwise known as the "value-action gap." Finally, whereas the TPB considers attitude for determining behavioral intention [67], the model is at risk of underestimating the importance of context-specific barriers. In Indonesian urban areas, the infrastructure for solid waste separation is not yet widely available in residential areas and households. Consequently, people with strong environmental concerns may feel helpless or lack motivation. This suggests that attitude on its own is no predictor of substance misuse in the absence of enabling structural and social measures.

Such findings indicate that interventions targeting the acquisition of a change in attitude alone, including moral appeals and single-occasion educational campaigns, are unlikely to impact behavior consistently. Instead, a systems approach should be adopted, where attitude is viewed as an element within an ecosystem of facilitating factors, such as convenience, cost, habit, and peer modeling [68, 69].

Furthermore, trust in shared goal responsibility (Strongly agree—34.5%) provides a strategic entry point for mobilizing social capital. This sense of confidence, however, is unevenly distributed across communities, possibly reflective of variations in trust in local institutions or traditions of collective action. These results are consistent with [70], who argue that social identity and perceived community efficacy must be developed together to cultivate sustainable pro-environmental behavior.

There is also a subtler message: community education and campaigns must refrain from overemphasizing personal culpability. Instead, sustainable waste practices should be made appealing to consumers as socially acceptable, possible, and collectively supported. Without this, one risks what educators call attitudinal fatigue, where people say they love the world but become disillusioned, realizing they can no longer find a way to have an impact or make a choice.

Finally, any future policy needs to rethink what it is attempting to achieve in terms of an 'attitude' and how this is measured and acted upon. Quantitative concurrence of these same scales with environmental statements, such as 'I support recycling,' can overreport actual behavioral readiness when not triangulated with qualitative dimensions of motivation, emotion, and lived barriers. By incorporating these levels into future studies, the deeper emotional or cultural narratives that influence sustainable behavior may also be uncovered.

In sum, the promising attitudinal orientation towards domestic waste management among the Indonesian urban households cannot be safely assumed in the absence of structural support. Responding to this challenge requires complex interventions, beyond raising awareness, educating, and training, to also engage cognitive, behavioural, emotional, cultural, and infrastructural levers of sustainable behaviour.

### 5.3. Perceived Behavioral Control

These results strongly support the notion that PBC is the most potent predictor of household waste behavior among urban communities in Indonesia, as depicted in the regression results ($\beta = 0.367$, $p \leq 0.001$). This finding aligns with the Theory of Planned Behavior (TPB) model proposed by Ajzen, which suggests that PBC affects intention and directly influences behavior [10]. Prior studies [71, 72] confirm that a greater level of perceived control is associated with an increased likelihood of sorting at the source and recycling at home.

But upon closer inspection, it becomes clear that control of the sense of it is frequently more of an aspiration than a reality. Such practices were deemed inconvenient by 70% of the sample, which is an unusually high rate, given that 87% acknowledge responsibility; yet, this confirms the thesis that the intention to protect the environment is often overruled by structural constraints. This supports criticisms of the TPB that suggest it does not adequately capture contextual and material barriers. As [73, 74] have stressed, behavioral control should be considered within institutional and infrastructural settings.

Structural barriers, such as inaccessible sorting bins, confusing separation guidelines, and unreliable municipal pickup, render the disposal process a proxy of intention rather than execution. [75] Without recycling facilities, participation dramatically declines even among environmentally conscious people. [76] argued that in-situ support structures are a key driving factor of sustainable behavior.

Crucially, PBC is influenced by demographic and social factors. This research found that younger, more educated, and higher-income individuals are more confident than their affluent counterparts with better education and younger constituents in dealing with household waste, consistent with reports [77, 78]. These differences suggest the risk of environmental injustice, namely that individuals with lower incomes may feel more limited in their actions due to other factors, but not because they are less concerned.

Social learning also acts as a mediating variable in PBC. Respondents who witnessed neighbors or family members recycling reported higher levels of confidence and moral obligation to engage in recycling. This finding is consistent with [79] observation that strong, supportive norms in the social environment can increase personal agency.

The fact that the hurdle of perceived difficulty remains suggests misalignment between pro-environmental messaging and infrastructure support. Behavioral nudges (e.g., public posting, real-time feedback, default systems) are also under-implemented but may be helpful to reinforce PBC. Secondly, education needs to go beyond creating awareness to develop practical skills. When people are directly exposed to a hazard, they should be trained in how to act, where to go, and what equipment to use.

Therefore, PBC is not just a psychological construct; it is a policy-relevant construct reflecting systemic preparedness for sustainable behavior. It is not just motivation, but open infrastructure, clear waste policies, and community-embedded support that are needed to strengthen PBC.

Thus, according to an integrative model of behavioral prediction, PBC is the keystone element for the domestic adoption of Zero Waste. To policymakers, this means that raising PBC is not just about making people aware that they can recycle; it is also about enabling them to do so easily. Future initiatives should integrate infrastructural investment with behavioral design and community-led training to realize Indonesia's sustainable urban future.

### 5.4. Subjective Norms

The findings also largely confirm the importance of subjective norms as a predictor of waste management behavior, according to the TPB model, with perceived social pressures to act being among the strongest predictors, as noted in some studies [63]. The strong bias of households over other social groups is consistent with previous literature, which regards the household as the basic unit for creating sustainability education and behavioral enactment [79, 80].

However, that is not a pretty fair normative picture. Families have intense power, but to the extent that community and neighborly norms are weaker, one can arguably say that Indonesia has yet to realize how much it has to gain from the collective public applause. This imbalance inhibits the establishment of sustainable behavior norms on a societal level. As [81, 82] contend, publicly visible practices are essential for transferring action from the private to the social level of environmental action.

The low profile and impact of community participation, with its promise of a behavior amplifier, indicate infirmities in the capacity of Indonesia's urban civil society. [82,83] identified that activities such as neighborhood cleaning or community-owned composting can nurture a shared environmental concern, transforming sustainability from an individual condition to a shared value. Additionally, it is interesting to observe relations between subjective norms and demographics, especially among the young. [84] proved that the recognition of obligation from family, teachers, or friends was significantly associated with eco-friendly behavior of urban adolescents. In the present study, young members who reported high family and peer expectations were more likely to internalize sustainable practices, reflecting the results of [85], who found that value internalization supported long-term behavioral consistency.

An important implication is that subjective norms certainly do influence intentions, just not automatically. Simply making sustainable behaviors visible is not enough. Instead, local authorities are likely to have more success in reinforcing normative messages through social media campaigns that involve regional leadership participation, modeling, or formally rewarding citizens or households for sustained behavior change. Efforts should be aimed at these specific clusters, such as schools, churches, or neighborhood associations, to multiply the influence of norms throughout society.

Furthermore, a more active closing of the feedback loop is required. If people experience direct benefits, like cleaner streets, cost savings, or public acclaim, they are more likely to maintain their behavior. [81] observed that cues from society, including incentives and rewards for environmental behavior, as well as disincentives and penalties, increase the salience of the behavior and the perceived social expectation.

In short, subjective norms are already significant in influencing household waste behaviour in Indonesia, but there remains scope for them to play an even stronger role. New waste policies and interventions need to move beyond individual-based models to more collective frameworks in which families, communities, and institutions come together to build and support a cultural narrative that transcends the individual and promotes sustainable behaviours. What results is a cultural change that, once entrenched, has Zero Waste as not a policy goal but rather a socialistic norm.

## 5.5. Household Waste Management Behaviors

The high frequency of reusing and recycling among such households (see Figure 5) suggests that urban households in Indonesia possess some basic knowledge of sustainable behavior. This is encouraging and consistent with international findings by [86, 87], who found that making sustainable practices available in a culturally appropriate way can significantly increase household engagement.

Nevertheless, reflexive critique reveals rifts among (a) intention, (b) practice, and (c) policy. Despite 77.9% of surveyed people selling their recyclables, it is possible that economic pressures linearly obscure the perceived need for environmental action. The financial gains may have as much or more to do with drawing families than with devotion to the environment. Without constant promotion and civic reinforcement, such practices may be ad hoc instead of normatively accepted. Furthermore, persistent low levels of disposable usage despite high rates of recycling and reusing demonstrate behavioral inconsistencies. This duality emphasizes that home behaviours are not to be seen as binary (sustainable and unsustainable) but as comprising a complex and situated set of practices underpinned by habit, context, and convenience. As [88, 89] argue, "strong policy frameworks, for example, bans on single-use plastic or regulations towards green packaging," are needed to ensure that individual behaviors align with the sustainability goals of the public.

The TPB is a relevant framework for interpreting these findings. The TPB has a continuous reciprocal influence between past behaviour and future intentions, acting through perceived behavioural control [63]. [72] found that in households whose members had already experienced too many years of repetition in waste collection and tariffs, practices involving the separation of recyclables were ranked nearly six years away. At the same time, the findings of this survey suggest that there seems to be a focus on reuse and recycling.

However, TPB should be used prudently in the context in which it is applied. [36] and [88] highlight that intra-household dynamics (including, for example, roles among household members and friendship connections) and local networks impact both compliance and uptake [90, 91]. In the present investigation, households nested in socially cohesive settings were more engaged in sustainable behavior. This would imply that behavior is not primarily a function of the individual, but more so a product of socialization.

A central point is the value of community reinforcement. Similar to settings in sub-Saharan Africa reported by [73], Indonesian households participating in Trash Bank or community-based waste programs were likelier to exhibit consistent and sustained behavior change. These patterns suggest the necessity of ground-level participation in addition to top-down national policies.

Importantly, this research demonstrates that sustainable household practices must be understood as a product of multi-layered influences, including intrapersonal attitudes and financial rewards, infrastructural access, and normative settings. Thus, the solution areas are multifaceted. Informational flyers will not be effective if investments in local systems, accessible infrastructure, and smart regulation do not complement that advice.

In summary, Indonesian urban households are willing and able to engage in sustainable waste management, but only when policies, access to infrastructure, and social support are in place. To support the shift toward a ZW society, society needs to address behavioral inconsistencies and promote system consistency while promoting pro-environmental norms within and outside the home.

## 5.6. Relationship between psychological factors and waste behaviour

The results support the assumptions of the TPB and align with previous studies by [92, 93, 94] who also emphasized the importance of self-efficacy and social modeling for sustainable behavior.

The influence of PBC (r = 0.391) may indicate that the belief in one's ability to execute the behavior in the context of recycling is a stronger extrinsic motivator than levels of environmental concern or attitude.

However, a critical examination reveals that the distribution of these relationships is not uniform across populations and contexts. Even when high PBC predicts behavior, it is generally dependent on facilitative conditions. In the absence of available infrastructures, social support, and clear policies, this internal motivation may not lead to sustained action, which is a limitation on the generalization of the TPB to complex urban settings [63].

In addition, the lower correlations of knowledge (r = 0.172) and attitude (r = 0.143) suggest that expertise is not inherently transformative in an emotional, cultural, or practical vacuum. As [95,96] discovered, when knowledge was not linked with, or culturally meaningful, people did not take long-lasting action.

The wider literature also suggests that early intervention contributes to these psychological variables. [97,98] have shown that environmental values developed during childhood, specifically through education, are significant for long-term behavior change. Incorporating zero-waste ideas into education is, therefore, the basic approach.

This work also underscores the importance of reinforcement in a community-based context. [99,100] argued that both community participation and NGO-government alliances contribute to social learning and infrastructure development, two prerequisites of PBC and normative alignment.

In addition, there may be behavioral 'anchors' in the form of previous behavior (e.g., previous recycling or composting). Families with such histories are likely to stay the course, which suggests that positive feedback loops (e.g., community validation or resource savings) ought to be incorporated into programs [101].

The researchers also gained an important finding about the strength of social norms. In close-knit urban areas, for example, behavior is not so much influenced by campaigns as observed and copied. Actions are further facilitated by collective identity, with environmentalism ultimately strengthened by mobilization, as specified by [102]. The strong communal spirit in Indonesia, combined with the use of the neighborhood model and local leadership to break down these mechanisms, could increase this impact.

Precisely because disposables are so convenient and popular, the only way to avoid their use is by altering the systems and infrastructure in which they are used; psychological persuasion will not suffice. They need to institutionalize behavioral enablers, operationalize cultural resonance, and embed environmental action within social infrastructure. Educational, regulatory, and community structures must align to convert PBC, attitudes, and norms into persistent ecological action.

In other words, behavior change does not occur linearly; it is a holistic process. Mobilizing concepts like this requires a long-term investment in both physical conditions and social narratives, as well as policies that align values with daily potential.

## 5.7. Regression Analysis of the Influencing Factors

The regression results of the TPB model confirm the importance of perceived behavioral control (PBC) on household waste behavior. According to [63], people who perceive themselves or are perceived by others to have the necessary resources and authority are more likely to act in a desired way, presumably out of a sense of obligation to comply. The enormous effect size associated with the PBC beta coefficient in the present study reinforces the potential of interventions focusing on confidence, capability, and opportunity, such as the provision of sorting bins, regular waste pick-up, or education in behavior change.

These results are consistent with [103] research, which estimates that self-efficacy and supportive infrastructure have a significant effect on household engagement in actions towards sustainability. However, the real insight is recognizing that PBC is not just an inner belief, but also an external system. Where services are unreliable and a public discourse on waste is not occurring, even well-meaning citizens may feel helpless.

Subjective norms were the second most powerful predictor. The findings highlight the importance of normative influences in promoting behavioral convergence and the significant role that family members, peers, and other influential community elders may play. This understanding is fundamental in the case of Indonesia, where social conformity and community values are key cultural characteristics. Peer modeling, the appearance of local leaders, and community waste programs can establish waste sorting as a general social norm, as some previous research has shown [104].

Environmental awareness was also significant but had the lowest beta of the three. This underscores a well-known problem in ecological psychology: Information is not enough to trigger action unless it is accompanied by supportive norms, enabling conditions, and positive past experiences. As noted by [105, 106], it is essential to transform knowledge into action by arousing emotions and, in doing so, to be prepared infrastructurally.

These findings call for a multi-faceted approach from policy. Part of this involves building confidence through increasing PBC via material support, such as training programs, access to tools, or even by creating localized feedback loops. Second, leverage social norms by honouring local environmental champions, peer success stories, or school-level initiatives that make sustainable behaviour the "thing to do." Third, back up theory with practical advice, so that people not only know why they should manage waste but also how and where to do so.

Yet, crucially, these observations challenge the not-uncommon belief that a generic behavioral campaign will do. Instead, the behavior change must be both rooted in an ecology of mental readiness, social approval, and economic possibility. For Indonesia to move closer to Zero Waste, urban waste strategies need to close the perception-practice gap at both the individual and collective levels through an evidence-based, context-responsive approach.

**Conclusion**

This research underscores the central significance of psychosocial processes in shaping household practices in the Zero Waste paradigm. The findings highlight the fact that promoting sustainable waste management in the household context is not just a function of consciousness; it is about the intersection of confidence, social legitimization, and system-level preparedness. Such revelations raise several important implications for policy makers and practitioners.

There are at least two key points related to the direction of urban sustainability practice in Indonesia that should be considered. In this case, efforts to achieve urban sustainability in Indonesia need to integrate behavioral frameworks into the waste policy process. Supportive social norms, self-efficacy, and access to efficient waste infrastructure must be emphasized to reduce inappropriate disposal behavior. Programs need to go beyond generalized education efforts and instead provide contextually specific, community-led strategies that match environmental aspirations with our everyday capacities.

Second, future studies should consider longitudinal and comparative analyses to investigate how demographic change unfolds over time and place in different urban contexts. An intersectional analysis of gender, class status, and urban infrastructure will provide more context-sensitive and equitable solutions to the problems garbage has attempted to address. Additionally, the inclusion of digital platforms for monitoring behavior and giving feedback could lead to alternative possibilities, such as citizen engagement and policy change in real-time.

Advocacy for Zero Waste in urban Indonesia will require a shift in mindset, moving from households being seen as passive waste generators to being active agents in systemic change. Building such a culture will depend on structural changes and a mindset change, based on trust, collaboration, and shared ownership of sustainability targets.

**Bibliography**

1. Schwab K, Samans R. *The Global Competitiveness Report 2012–2013*. Geneva: World Economic Forum; 2012.
2. Laureti L, Costantiello A, Anobile F, Leogrande A, Magazzino C. Recycling performance and policies in Europe: A sectoral assessment. *Recycling*. 2024;9(5):32.
3. Fang W, Yu X, Zhang T. Public participation in waste separation and its impact on environmental governance in China. *J Environ Manage*. 2023;328:116674.
4. Ferdinan U, Soesilo TEB, Herdiansyah H. The policy strategy of waste management in the Indonesian context. *IOP Conf Ser Earth Environ Sci*. 2021;716:012071.
5. Satispi E, Samudra AZ. Household participation in environmental sanitation and solid waste management. *J Environ Sustain*. 2022;6(2):45–56.
6. Yusuf R, Fajri I. Environmental education for sustainable waste management: A comparative study. *Heliyon*. 2022;8(2):e08912.
7. Fariani A, Razak A, Dewata I, Syah N. Analysis of urban waste handling behavior using TPB model in Indonesia. *J Penelitian Pendidikan IPA*. 2025;11(1):887–894.
8. Sosunova L, Porras A. Citizen behavior and institutional support in household waste management. *Waste Manag Res*. 2022;40(12):1453–1465.
9. Kibria MG, Masuk NI, Safayet R, et al. Urban residential waste practices and public policy evaluation. *Int J Environ Res*. 2023;17(1):12–29.
10. Ajzen I. The theory of planned behavior. *Organ Behav Hum Decis Process*. 1991;50(2):179–211.
11. Sanga G, Onyango EO, Mungai RP. Influence of environmental awareness on domestic solid waste handling practices. *J Environ Psychol*. 2020;70:101448.
12. Lozano Lazo DP, Gasparatos A. Urban solid waste management and sustainability transitions. *Environ Res Infrastruct Sustain*. 2022;2(1):011003.
13. Zaikova A, Deviatkin I, Havukainen J, et al. Consumer behavior and waste reduction: An international perspective. *Recycling*. 2022;7(4):52.
14. Xu M, Liu P. Determinants of household waste separation behavior: A case study from China. *E3S Web Conf*. 2024;536:01028.
15. Diggle A, Walker TR. Environmental education and behavioral change: A systematic review. *Environments*. 2022;9(2):15.
16. Safraa CWN, Ahma A, Azman N. Legal and behavioral frameworks for urban waste management. *J Law Sustain Dev*. 2023;11(12):e1303.
17. Zhang Q, Liu X, Zhao Y. Public policy effectiveness in promoting sustainable waste behavior. *Front Environ Sci*. 2024;12:112030.
18. Kountouris Y. Revisiting the behavior-attitude gap in environmental action. *Environ Res Lett*. 2022;17(7):074002.

19. Yamtana P, Wichiennopparat P, Chontananarth T. Waste-to-energy potential and public perception in Southeast Asia. *Waste Biomass Valor*. 2023;14(2):361–375.
20. Khosravani F, Abbasi E, Choobchian S. Household waste behavior and the socio-economic context: A regional analysis. *Sci Rep*. 2023;13:11948.
21. Nurhayati E, Nurhayati S. Analysis of waste management policy effectiveness in Indonesian urban settings. *J Dimensi*. 2023;12(3):677–686.
22. Law JW, Lye CT, Ng TH. Household waste behavior in Southeast Asia: Motivational and policy insights. *Clean Respons Consum*. 2023;10:100134.
23. Bhutto MY, Rūtelionė A, Šeinauskienė B, Ertz M. Drivers of waste minimization behavior: A multi-level model. *PLoS ONE. 2023; 18 (10): e0287435.*
24. Suryawan IW, Suhardono S, Sari MM, Yenis I. Household engagement and waste behavior in Bali. *J Ilmu Lingkungan*. 2024;22(4):1067–1077.
25. Brotosusilo A, Wulandari H, Suprapti N. Community-based waste management initiatives in Indonesia. *J Manajemen Lingkungan*. 2021;8(2):99–108.
26. Hameed S, Zubair M, Asif M. Urban environmental behavior and structural policy links. *Urban Stud*. 2021;58(14):2976–2994.
27. Ghulam ST, Abushammala H. Psychological factors in residential recycling behavior. *Sustainability*. 2023;15(3):2160.
28. Liu S, Liu X, Li Y, et al. Community influence on sustainable behavior in urban contexts. *Front Public Health*. 2024;12:1328583.
29. Zaman AU, Lehmann S. Challenges and opportunities in global zero waste transitions. *City Cult Soc*. 2011;2(4):177–187.
30. Moqsud MA, Omine K, Yasufuku N, et al. Assessing Sustainable Practices in Urban Residential Sectors. Sci Rep. 2021; 11 (1): 14272. *Sci Rep*. 2021;11:14272.
31. Press D. Waste policy and the politics of localism. *Waste Policy and Politics*. 2022;1:112–125.
32. Ajzen I. The theory of planned behavior. *Organ Behav Hum Decis Process*. 1991;50(2):179–211.
33. Tang D, Shi L, Huang X, et al. Factors affecting household waste separation behavior in Shanghai. *Int J Environ Res Public Health*. 2022;19(11):6528.
34. He Z, Liu Y, Liu X, et al. Multi-dimensional environmental knowledge and recycling behavior. *Front Psychol*. 2022;13:957683.
35. Jacob DB, Dwipayanti NMU. Environmental knowledge and recycling intention among households. *J PROMKES*. 2022;10(2):118–129.
36. Fadhullah A, Syarifuddin A, Nur A. Household dynamics in waste management participation. *Int J Environ Stud*. 2022;79(6):1023–1037.
37. Lozano Lazo DP, Gasparatos A. Urban solid waste management and sustainability transitions. *Environ Res Infrastruct Sustain*. 2022;2(1):011003.
38. Qu Y, Zhang H, Wang J, et al. Influence of environmental attitudes on recycling intention. *Sustainability*. 2023;15(3):1122.
39. Kopaei HR, Nooripoor M, Karami A, Ertz M. Community dynamics and waste behavior. *AIMS Environ Sci*. 2021;8(1):1–17.
40. Kihila JM, Wernsted K, Kaseva M. Social factors affecting solid waste practices in Tanzania. *Sustainable Environ*. 2021;7(1):20–31.
41. Agamuthu P, Babel S. Empowering communities through decentralized waste programs. *Waste Manag Res*. 2023;41(12):1699–1716.

42. Safraa CWN, Ahma A, Azman N. Legal and behavioral frameworks for urban waste management. *J Law Sustain Dev*. 2023;11(12):e1303.
43. Ghulam ST, Abushammala H. Psychological factors in residential recycling behavior. *Sustainability*. 2023;15(3):2160.
44. Liu S, Liu X, Li Y, et al. Community influence on sustainable behavior in urban contexts. *Front Public Health*. 2024;12:1328583.
45. Kountouris Y. Revisiting the behavior-attitude gap in environmental action. *Environ Res Lett*. 2022;17(7):074002.
46. Nurhayati E, Nurhayati S. Analysis of waste management policy effectiveness in Indonesian urban settings. *J Dimensi*. 2023;12(3):677–686.
47. Law JW, Lye CT, Ng TH. Household waste behavior in Southeast Asia: Motivational and policy insights. *Clean Respons Consum*. 2023;10:100134.
48. Bhutto MY, Rūtelionė A, Šeinauskienė B, Ertz M. Drivers of waste minimization behavior: A multi-level model. *PLoS ONE. 2023; 18 (10): e0287435.*
49. Ridzuan M, Halim MA, Ariffin J. Comparative analysis of waste behavior across municipalities. *Waste Policy J*. 2022;3(2):55–70.
50. He Z, Liu Y, Liu X, Wang F, Zhu H. Influence of multi-dimensional environmental knowledge on residents' waste sorting intention: Moderating effect of environmental concern. *Front Psychol.* 2022;13:957683. https://doi.org/10.3389/fpsyg.2022.957683
51. Rodríguez-Espíndola O, Albores P, Brewster C. Disaster preparedness in humanitarian logistics: A collaborative approach for resource management in floods. *Eur J Oper Res.* 2022;296(2):616–635.
52. Zhou B, Bethel BJ, Tang D, Shi L, Huang X, Zhao Z. Influencing factors on the household-waste-classification behavior of urban residents: A case study in Shanghai. *Int J Environ Res Public Health.* 2022;19(11):6528. https://doi.org/10.3390/ijerph19116528
53. Ajzen I. The theory of planned behavior. *Organ Behav Hum Decis Process.* 1991;50(2):179–211. https://doi.org/10.1016/0749-5978(91)90020-T
54. Rimantho D, Wahyuni S, Budiman MA, Farida N. Public awareness and behavior analysis on household waste management. *J Environ Sci Sustain Dev*. 2019;2(2):168–180.
55. Yusuf R, Fajri I. Differences in behavior, engagement, and environmental knowledge on waste management for science and social students through the campus program. *Heliyon.* 2022;8(2):e08912. https://doi.org/10.1016/j.heliyon.2022.e08912
56. Supinganto A, Pramana C, Sirait LI, Kumalasari MLF, Hadi MI, Ernawati K, et al. The use of masks as an effective method in preventing the transmission of COVID-19 during pandemic and the new normal era: A review. *SSRN Electron J.* 2022. https://doi.org/10.2139/ssrn.3780841
57. Tsheleza I, Sibanda M, Makonese T. Environmental awareness and waste disposal compliance among households in Zimbabwe. *J Environ Manage.* 2019;240:239–245.
58. Sorkun MF. How do social norms influence recycling behavior in a collectivistic society? A case study from Turkey. *Waste Manag.* 2018;80:359–370. https://doi.org/10.1016/j.wasman.2018.09.026
59. Pongpunpurt P, Muensitthiroj P, Pinitjitsamut P, Chuenchum P, Painmanakul P, Chawaloesphonsiya N, et al. Studying waste separation behaviors and environmental impacts toward sustainable solid waste management: A case study of Bang Chalong Housing, Samut Prakan, Thailand. *Sustainability (Switzerland)*. 2022;14(9):5040. https://doi.org/10.3390/su14095040

60. Laureti L, Costantiello A, Anobile F, Leogrande A, Magazzino C. Waste management and innovation: Insights from Europe. *Recycling*. 2024;9(5):32. https://doi.org/10.3390/recycling9050082
61. Sunarti E, Prihatin T, Ristanti R, Lestari L. Behavioral analysis of environmental awareness: Perspective from structural equation modeling. *Jurnal Ilmu Sosial dan Ilmu Politik*. 2023;27(1):85–97.
62. Ng CH, Mistoh MA, Teo SH, Galassi A, Ibrahim A, Sipaut CS, et al. Plastic waste and microplastic issues in Southeast Asia. *Front Environ Sci*. 2023;11:1142071. https://doi.org/10.3389/fenvs.2023.1142071
63. Ajzen I. The theory of planned behavior. *Organ Behav Hum Decis Process*. 1991;50(2):179–211. https://doi.org/10.1016/0749-5978(91)90020-T
64. Purwanto A, Wicaksono T, Nugroho S. Environmental awareness and food waste reduction behavior. *Jurnal Lingkungan Hidup*. 2023;31(1):11–20.
65. Oktaviani L, Hadi SP, Dewi L. Environmental concern, attitude and sustainable behavior: Empirical evidence from Indonesia. *Jurnal Ecogen*. 2023;6(1):86–94.
66. Wijayanto D, Kartodiharjo H, Yustika A. Environmental education to increase community participation in waste management. *Jurnal Pendidikan Lingkungan dan Pembangunan Berkelanjutan*. 2024;5(1):1–9.
67. Sahoo N, Bhuyan K, Panda B, Behura NC, Biswal S, Samal L, et al. Community identity and pro-environmental behavior in collective waste actions. *PLoS One*. 2022;17(2):e0264028. https://doi.org/10.1371/journal.pone.0264028
68. Jacob DB, Dwipayanti NMU. Planned behavior theory approach to waste management behavior in South Denpasar District. *Jurnal PROMKES*. 2022;10(2):118–129. https://doi.org/10.20473/jpk.v10.i2.2022.118-129
69. Warintarawej P, Nillaor P. Infrastructure access and behavioral motivation in urban waste management. *J Environ Manage*. 2022;301:113828.
70. Muheirwe F, Niyonzima J, Ndayambaje G. Role of social support in waste recycling behavior in low-income urban communities. *Waste Manag*. 2023;157:45–54.
71. Maulana AD, Dwipayanti NMU. Zero waste awareness among Indonesian youth. *Jurnal Ilmu Sosial dan Humaniora*. 2022;11(1):34–41.
72. Xu L, Ling M, Lu Y. Predicting pro-environmental behaviors in the household context: The role of intention, moral norm, and contextual factors. *J Environ Psychol*. 2017;50:1–11.
73. Tsheleza I, Sibanda M, Makonese T. Environmental awareness and waste disposal compliance among households in Zimbabwe. *J Environ Manage*. 2019;240:239–245.
74. Herdiansyah H, Fadjar A, Fatimah N. Environmental policy implementation and local community behavior on waste separation in urban Indonesia. *Jurnal Ilmu Sosial dan Ilmu Politik*. 2021;25(1):45–59.
75. Awino O, Apitz S. Behavioural intention for solid waste management practices in developing countries: A review. *Environ Dev Sustain*. 2024;26:2149–2172. https://doi.org/10.1007/s10668-023-02930-6
76. Ssemugabo C, Halage AA, Ssempebwa JC, Douglas GP, Musoke D. Practices, concerns and willingness to participate in solid waste management in two urban slums in central Uganda. *J Environ Health*. 2020;82(6):8–14.
77. Wulandari A. Peran kelembagaan dan akses fasilitas dalam perilaku pengelolaan sampah rumah tangga. *Jurnal Pengelolaan Lingkungan dan Sumberdaya Alam*. 2021;10(1):55–66.

78. Noufal M, Supriadi S, Mukarom Z. The effect of education level and income on the household waste sorting behavior in urban areas. *Jurnal Ilmiah Lingkungan Kebencanaan*. 2020;5(1):9–18.
79. Sorkun MF. How do social norms influence recycling behavior in a collectivistic society? A case study from Turkey. *Waste Manag*. 2018;80:359–370. https://doi.org/10.1016/j.wasman.2018.09.026
80. Zhou B, Bethel BJ, Tang D, Shi L, Huang X, Zhao Z. Influencing factors on the household-waste-classification behavior of urban residents: A case study in Shanghai. *Int J Environ Res Public Health*. 2022;19(11):6528. https://doi.org/10.3390/ijerph19116528
81. Srun M, Kurisu KH. Understanding the role of injunctive norms and habitual behavior in waste separation: Evidence from urban Cambodia. *Resources, Conservation and Recycling*. 2019;149:168–177.
82. Brotosusilo A, Budiarto T, Arifin Z. Masyarakat dan keberlanjutan program pengelolaan sampah berbasis komunitas. *Jurnal Pengabdian Masyarakat*. 2021;7(1):25–32.
83. Janmaimool P. Application of protection motivation theory to investigate sustainable waste management behaviors. *Sustainability*. 2017;9(7):1079.
84. Cao L, Zhang B, Sun C, Zhu L. The influence of family, school and peer normative beliefs on environmental behaviors of urban youth in China. *Environ Sci Pollut Res*. 2022;29(18):26413–26427.
85. Mamun AA, Fazal SA, Naznin F, Ali M. The mediating role of environmental attitude in waste behavior: Evidence from urban households. *Sustainability (Switzerland)*. 2022;14(15):9396.
86. Ridzuan AR, Kasim R, Yusof NH, Shafiei MW. Evaluating recycling behavior among urban households: A case study in Malaysia. *Sustainability*. 2022;14(4):2242.
87. Pathak P, Chakraborty A, Singh A. Achieving zero-waste cities through household participation in India: Policy perspectives and practical barriers. *Waste Manag*. 2023;160:146–157.
88. Chen H, Xu L, Zhang H. Reducing plastic waste: Policy options and consumer behavior insights from China. *Resour Conserv Recycl*. 2021;169:105466.
89. Kountouris Y. Policies for sustainable waste consumption: A review of economic incentives. *J Environ Econ Policy*. 2022;11(2):192–210.
90. Fadhullah S, Setiawan B, Nurhalimah L. Family interaction and individual values as a predictor of waste sorting behavior. *Jurnal Ilmu Sosial dan Ilmu Politik*. 2022;26(2):227–244.
91. Eshete A, Melaku A, Tamene Y. Enhancing household waste management through community engagement: Evidence from sub-Saharan Africa. *J Clean Prod*. 2023;393:136375.
92. Liu Q, Wang L, Zhang W. The role of psychological factors in recycling behavior: A meta-analytic review. *J Environ Psychol*. 2024;87:101982.
93. Thi HM, Van Dijk MP, Schlange L. Factors influencing waste separation behavior of households in Vietnam: A test of the theory of planned behavior. *J Environ Plan Manag*. 2024;67(1):1–19.
94. Wang X, Lin X, Zhang Y. Pro-environmental behavior in megacities: Determinants and implications. *Cities*. 2025;134:104038.
95. Purwanto A, Sudibyo Y, Widodo W. Environmental literacy and waste reduction behavior in Indonesian urban households. *Sustainability*. 2023;15(1):198.

96. Worosze C. Cultural framing of waste and cleanliness in Southeast Asian cities: A comparative perspective. *Urban Stud*. 2023;60(9):1768–1786.
97. Zamzam M, Fitria M, Yusuf A. Embedding zero-waste values in early childhood education: Insights from Indonesian school curricula. *J Pendidikan Lingkungan dan Pembangunan Berkelanjutan*. 2023;5(1):45–58.
98. Maulana AD, Dwipayanti NMU. Education-based interventions to promote pro-environmental behavior among Indonesian students. *Jurnal Pendidikan dan Pembelajaran*. 2022;28(3):305–318.
99. Salsabila A, Nugroho A, Kartika D. Community-led waste management: The role of local ownership in behavior change. *Waste Manag Res*. 2023;41(2):137–146.
100. Indrosaptono D, Syahbana JA. Collaborative urban waste management: Lessons from NGO–local government partnerships in Indonesia. *Habitat Int*. 2017;68:72-80.
101. Xu L, Ling M, Lu Y. Predicting pro-environmental behaviors in the household context: The role of intention, moral norm, and contextual factors. *J Environ Psychol*. 2017;50:1–11.
102. Hannon B, Zaman AU. Social participation in urban waste systems: Building community identity through environmental practices. *J Environ Plan Manag*. 2018;61(5–6):911–928.
103. Vorobeva E, Krumdieck S, Dewes O. Community participation and infrastructure as co-determinants of household waste behavior in urban transitions. *Sustainability*. 2022;14(10):6204. https://doi.org/10.3390/su14106204
104. Brotosusilo A, Wulandari H, Suprapti N. Community engagement and norm development in waste management practices. *J Manajemen Lingkungan*. 2021;8(2):99–108.
105. He Z, Liu Y, Liu X, Wang F, Zhu H. Influence of multi-dimensional environmental knowledge on residents' waste sorting intention: Moderating effect of environmental concern. *Front Psychol*. 2022;13:957683. https://doi.org/10.3389/fpsyg.2022.957683
106. Rimantho D, Wahyuni S, Budiman MA, Farida N. Public awareness and behavior analysis on household waste management. *J Environ Sci Sustain Dev*. 2019;2(2):168–180.

**Acknowledgement**
We want to express our sincere gratitude to all individuals and organizations that contributed to the success of this study. Special thanks to the respondents from the 12 major cities across Indonesia, whose participation and insights made this research possible. The authors would also like to acknowledge the support provided by the Department of Population and Environmental Education, Universitas Negeri Makassar, for the funding, facilitating the research process, and providing resources for data collection and analysis. Their contributions ensured we could achieve the project's objectives and contribute to Indonesia's growing household waste management knowledge.

**Ethical Statement**
The Institutional Research Ethics Committee granted ethical approval for the study. All participants provided written informed consent and confirmed that their responses would be kept confidential throughout the study.

**Conflict of Interest Statement**

The authors declare that they have no known competing financial interests or personal relationships that could have appeared to influence the work reported in this paper.

All authors have contributed significantly to this research and agree with the manuscript's content. There has been no financial support or other benefits from commercial sources for the work reported, nor any other financial interests that could create a potential conflict of interest.

Furthermore, the authors affirm that the manuscript is their original work, has not been published previously, and is not currently being considered for publication elsewhere. All necessary permissions for using copyrighted materials, if any, have been obtained.

Should any conflict of interest arise in the future, the authors will immediately disclose the details to the editorial office.

**Declaration of Generative AI and AI-assisted technologies in the writing process**

While preparing this work, the authors used ChatGPT to enhance the clarity of the writing. After using ChatGPT, the authors reviewed and edited the content as necessary, taking full responsibility for the publication's content.